\pdfoutput=1

\documentclass[11pt,twoside,a4paper,cmspaper,final,collab]{cms-tdr}
\def\svnVersion{471377P}\def\cmsCernNoTag{CERN-EP-2018-091}\def\cmsCernDate{\today}\def\cmsMessage{Published in Physical Review Letters as \href{http://dx.doi.org/10.1103/PhysRevLett.121.062002}{\doi{10.1103/PhysRevLett.121.062002.}}}

\begin{document}\cmsNoteHeader{HIN-16-003}

\hyphenation{had-ron-i-za-tion}
\hyphenation{cal-or-i-me-ter}
\hyphenation{de-vices}
\RCS$Revision: 469262 $
\RCS$HeadURL: svn+ssh://svn.cern.ch/reps/tdr2/papers/HIN-16-003/trunk/HIN-16-003.tex $
\RCS$Id: HIN-16-003.tex 469262 2018-07-20 12:45:30Z mironov $

\ifthenelse{\boolean{cms@external}}{\providecommand{\suppMaterialRef}{\cite{suppMat}}}{\providecommand{\suppMaterialRef}{\nocite{suppMat}}}

\ifthenelse{\boolean{cms@external}}{\providecommand{\suppMaterial}{the supplemental material~\suppMaterialRef }}{\providecommand{\suppMaterial}{Appendix~\ref{app:suppMat}\suppMaterialRef}}

\newlength\cmsFigOneWidth
\ifthenelse{\boolean{cms@external}}{\setlength\cmsFigOneWidth{0.90\columnwidth}}{\setlength\cmsFigOneWidth{0.5\textwidth}}
\newlength\cmsFigTwoWidth
\ifthenelse{\boolean{cms@external}}{\setlength\cmsFigTwoWidth{0.90\textwidth}}{\setlength\cmsFigTwoWidth{0.90\textwidth}}
\ifthenelse{\boolean{cms@external}}{\providecommand{\cmsLeft}{top\xspace}}{\providecommand{\cmsLeft}{left\xspace}}
\ifthenelse{\boolean{cms@external}}{\providecommand{\cmsRight}{bottom\xspace}}{\providecommand{\cmsRight}{right\xspace}}

\newcommand	{\pythiahijing}{\PYTHIA{}+\HIJING\xspace}

\newcommand {\PbPb}  {\ensuremath{\mathrm{PbPb}}\xspace}
\newcommand {\pPb}  {\ensuremath{\Pp\mathrm{Pb}}\xspace}
\newcommand{\Pb}{\ensuremath{\mathrm{Pb}}\xspace}
\newcommand{\pp}{\ensuremath{\Pp\Pp}\xspace}

\newcommand{\ptave}{\ensuremath{\pt^{\text{ave}}}\xspace}
\newcommand{\etadijet}{\ensuremath{\eta_{\text{dijet}}}\xspace}
\newcommand {\ptone}{\ensuremath{{\pt}_{,1}}\xspace}
\newcommand {\pttwo}{\ensuremath{{\pt}_{,2}}\xspace}
\newcommand {\qq}{\ensuremath{Q^{2}}\xspace}

\cmsNoteHeader{HIN-16-003}
\title{\texorpdfstring{Constraining gluon distributions in nuclei using dijets in proton-proton and proton-lead collisions at $\sqrtsNN = 5.02\TeV$}{Constraining gluon distributions in nuclei using dijets in proton-proton and proton-lead collisions at sqrt(s[NN])=5.02 GeV}}

\date{\today}

\abstract{
The pseudorapidity distributions of dijets as functions of their average transverse momentum (\ptave) are measured in proton-lead (\pPb) and proton-proton (\pp) collisions. The data samples were collected by the CMS experiment at the CERN LHC, at a nucleon-nucleon center-of-mass energy of 5.02\TeV. A significant modification of the \pPb spectra with respect to the \pp spectra is observed in all \ptave intervals investigated. The ratios of the \pPb and \pp distributions are compared to next-to-leading order perturbative quantum chromodynamics calculations with unbound nucleon and nuclear parton distribution functions (PDFs). These results give the first evidence that the gluon PDF at large Bjorken $x$ in lead ions is strongly suppressed with respect to the PDF in unbound nucleons.
}

\hypersetup{%
pdfauthor={CMS Collaboration},%
pdftitle={Constraining gluon distributions in nuclei using dijets in proton-proton and proton-lead collisions at sqrt(sNN)= 5.02 TeV},%
pdfsubject={CMS},%
pdfkeywords={CMS, physics, heavy ion, jet, nuclear parton distribution function}
}

\maketitle

Relativistic heavy ion collisions aim to study the properties of the quark-gluon plasma (QGP)~\cite{Collins:1974ky,Cabibbo:1975ig,Freedman:1976xs,Shuryak:1977ut}, a deconfined state of quarks and gluons expected by quantum chromodynamics (QCD)~\cite{Karsch:2003jg} to exist for very high temperatures and energy densities. These studies are often performed by investigating changes in observables, such as jets and hadron spectra, in going from proton-proton (\pp) to heavy-ion collisions. The changes can be attributed to both initial-state effects (\eg, different parton distribution functions (PDFs) for heavy nuclei than for nucleons) and final-state effects due to the creation of the QGP~\cite{Salgado:2011wc,Albacete:2013ei}. Knowledge of the nuclear parton distribution functions (nPDFs) in heavy nuclei is thus crucial in extracting QGP properties from experimental data. While the quark nPDF of lead ions (\Pb) is well understood from deep inelastic scattering data~\cite{Arneodo:325160}, the gluon nPDF, which is particularly important for perturbative QCD (pQCD) calculations at the CERN LHC energies, is not well constrained. This is because of the limited amount of experimental data that are sensitive to the nPDFs of gluons at perturbative scales~\cite{Paukkunen:2014nqa,Paukkunen:2017bbm}. The pion spectra in deuteron-gold collisions at a nucleon-nucleon center-of-mass energy $\sqrtsNN = 200\GeV$~\cite{Adler:2006wg,Adams:2006nd} are the only relativistic heavy ion collision data from the BNL RHIC used in the global fits of nPDFs. Global fits including these data predict, with respect to the unbound PDFs, a suppression in the small Bjorken $x$ region $x \la 10^{-2}$ (\ie, shadowing), an enhancement in the intermediate $x$ region $10^{-2}\la{\times}\la 10^{-1}$ (\ie, antishadowing), and a suppression in the $x \ga 10^{-1}$ region (\ie, the EMC effect, named after its first observation by the European Muon Collaboration~\cite{Aubert:1983xm}) for gluon PDFs, as parametrized in the EPS09~\cite{Eskola:2009uj}, nCTEQ15~\cite{Kovarik:2015cma}, and EPPS16~\cite{Eskola:2016oht} nPDFs. These three nPDFs are similar in that they are all based on next-to-leading (NLO) perturbative QCD calculations. They differ in their parametrization of the three mentioned nuclear effects and in the input experimental data used in the global fit, with the EPPS16 nPDF being the only one using LHC dijet and \PW\ and \PZ\ bosons data from proton--lead (\pPb) collisions at $\sqrtsNN = 5.02\TeV$. However, the RHIC pion data can also be interpreted as nuclear modification of the parton-to-pion fragmentation function~\cite{Sassot:2009sh} without significant nuclear modification of the gluon PDF. This is the approach adopted in the deFlorian-Sassot-Stratmann-Zurita (DSSZ) nPDF~\cite{deFlorian:2011fp}. Therefore, high transverse momentum (\pt) jet data, which are relatively insensitive to a possible modification of the parton fragmentation, the underlying event (UE), and hadronization effects~\cite{Dasgupta:2007wa}, can provide crucial inputs to global fits of the nPDFs and thereby test their underlying parametrization assumptions. At the CERN LHC, jet measurements at high \pt can also be used to test the collinear factorization theorem in QCD~\cite{Collins:1989gx}, namely that the cross section of a process is a convolution in partonic momentum space of a perturbatively calculable part, with non-perturbative distributions of partons inside the hadrons involved in the process. The jet measurements are complementary to measurements of the \PJGy\ meson production cross-section in ultra-peripheral lead-lead (PbPb) collisions~\cite{Abelev:2012ba,Khachatryan:2016qhq} and low-\pt hadron spectra in \pPb collisions~\cite{ALICE:2012mj,Aad:2016zif,Khachatryan:2016odn}, which are sensitive to nPDFs at lower values of the momentum transfer in a collision (\qq).

The production of dijets, pairs of two jets consisting of the most energetic (leading) and the second most energetic (subleading) jets in the event, has been previously measured in \pPb collisions at the LHC~\cite{Chatrchyan:2014hqa,Adam:2015xea}. In contrast to what is observed in head-on \PbPb collisions, where QGP-induced gluon emission in the final state significantly alters the energy balance between the two highest-\pt jets~\cite{Chatrchyan:2011sx,Aad:2010bu,Chatrchyan:2012nia,Khachatryan:2015lha}, no significant dijet \pt imbalance is observed in \pPb data with respect to simulated \pp distributions~\cite{Chatrchyan:2014hqa}. Moreover, measurements of inclusive jet~\cite{ATLAS:2014cpa,Adam:2015hoa,Khachatryan:2015sva,Khachatryan:2016xdg} and inclusive charged-particle \pt spectra~\cite{Abelev:2014dsa,Khachatryan:2015xaa,Khachatryan:2015uja} also show no sign of modification at high \pt compared to \pp data. The relatively small or negligible final-state effects in \pPb collisions support the idea of using jets as probes for the nuclear PDF studies. Recent theoretical calculations suggest that the dijet pseudorapidity ($\etadijet = (\eta_{1} + \eta_{2})/2$) distribution in \pPb collisions provides strong constraints on the gluon nPDFs~\cite{Eskola:2013aya,Paukkunen:2014pha, Armesto:2015lrg,Khanpour:2016pph} because of the small experimental and theoretical uncertainties~\cite{Eskola:2013aya}. The measurement of the corresponding \pp reference spectra can further reduce the theoretical uncertainties for the extraction of the nPDFs~\cite{Armesto:2015lrg}.

In this Letter, measurements of dijet production are performed in \pPb and \pp collisions at $\sqrtsNN = 5.02\TeV$ recorded with the CMS detector and corresponding to integrated luminosities of $35\pm 1\nbinv$~\cite{CMS-PAS-LUM-13-002} and $27.4\pm 0.6\pbinv$~\cite{CMS-PAS-LUM-16-001}, respectively. To test the theoretical foundation of global analysis of nPDFs on collinear factorization, the ratios of the normalized \pPb and \pp \etadijet distributions ($\Pb/\pp$) are studied as a function of the dijet average transverse momentum ($\ptave=(\ptone+\pttwo)/2 \sim Q$) and compared with next-to-leading order (NLO) pQCD calculations involving different \qq values.

A detailed description of the CMS experiment can be found in Ref.~\cite{Chatrchyan:2008aa}. The silicon tracker, submerged in the 3.8\unit{T} magnetic field of the superconducting solenoid, is used to measure charged particles within the range $\abs{\eta}<2.5$. Also located inside the solenoid are an electromagnetic calorimeter (ECAL) and a hadron calorimeter (HCAL). The ECAL consists of more than 75,000 lead tungstate crystals, arranged in a quasi-projective geometry, and distributed in the barrel region ($\abs{\eta} < 1.48$) and in the two endcaps that extend up to $\abs{\eta} = 3.0$. The HCAL barrel and endcaps are sampling calorimeters composed of brass and scintillator plates, covering $\abs{\eta} < 3.0$. Iron hadron forward (HF) calorimeters, with quartz fibers read out by photomultipliers, extend the calorimeter coverage up to $\abs{\eta} = 5.2$. A muon system located outside the solenoid and embedded in the steel flux-return yoke is used for the reconstruction and identification of muons up to $\abs{\eta} = 2.4$. The detailed Monte Carlo (MC) simulation of the CMS detector response is based on \GEANTfour~\cite{Agostinelli:2002hh}.

The event samples are selected online with dedicated triggers requiring at least one jet with $\pt > 40$, 60, or 80\GeV, and are filtered offline to reject the beam-gas interaction induced background events~\cite{Chatrchyan:2014hqa}. In addition, \pPb collisions are selected by requiring a coincidence of at least one of the HF calorimeter towers, with more than 3\GeV total energy, from the HF detectors on both sides of the interaction point. Events are also required to have at least one reconstructed primary vertex with two or more associated tracks. This vertex is required to have a distance from the nominal interaction point of less than 15 and 0.15\unit{cm} in the  longitudinal (along the beam axis) direction and in the transverse plane (perpendicular the beam axis), respectively. In the \pPb data sample, there is a 3\% probability to have at least one additional interaction in the same bunch crossing (pileup). A pileup filter is employed~\cite{Chatrchyan:2013nka}, which rejects more than 90\% of the pileup events and removes 0.01\% of the events without pileup. The filter uses the longitudinal and transverse distance between the leading vertex (the vertex with the highest number of associated tracks) and the vertex with the second largest number of associated tracks as criteria for identifying and removing pileup events. In the \pp analysis, the pileup rejection procedure is not applied because of the significantly lower pileup rate (about a factor of three compared to \pPb).

Offline, jet reconstruction is performed using the CMS particle-flow (PF) algorithm~\cite{Sirunyan:2017ulk}. By combining information from all subdetector systems, the PF algorithm attempts to identify all stable particles in an event, classifying them as electrons, muons, photons, as well as charged and neutral hadrons. Jets are reconstructed from these PF candidates using the anti-\kt sequential recombination algorithm~\cite{Cacciari:2008gp,Cacciari:2011ma} with a distance parameter $R=0.3$, as implemented in the \FASTJET\ package~\cite{Cacciari:2011ma}. The reconstructed jets are then calibrated following the steps described in Refs.~\cite{Chatrchyan:2011ds,Khachatryan:2016kdb}.

Jets with pseudorapidity in the laboratory frame $\abs{\eta_{\text{lab}}}<3.0$ are used in the final \pPb analysis. Because of the different energies of the proton (4\TeV) and lead (1.58\TeV per nucleon) beams, the nucleon-nucleon center-of-mass frame is boosted in the detector frame. During part of the data-taking period, the directions of the proton and lead beams were reversed. For the data set taken with the opposite-direction proton beam, the sign of the standard CMS definition of $\eta$ was flipped so that the proton beam always moves towards positive $\eta$. Therefore, a massless particle emitted at $\eta_{\mathrm{cm}}=0$ in the nucleon-nucleon center-of-mass frame will be detected at $\eta_{\text{lab}}=+0.465$ in the laboratory frame. As described above, data from \pPb collisions are measured and presented in a symmetric region around $\eta=0$ in the laboratory frame. In order to obtain \pp data over the same $\eta$ range in the nucleon-nucleon center-of-mass frame, jets in the interval $-3.465 < \eta < 2.535$ are used. When studying \pp and \pPb data together, and also for the purposes of presentation, \etadijet for \pp data is shifted by $+0.465$, so that both sets of data span $\abs{\etadijet} < 3.0$ in the center-of-mass frame.

This analysis is carried out using events required to have a dijet with a leading jet of $\ptone > 90\GeV$, a subleading jet of $\pttwo>20\GeV$, and $\Delta \phi_{1,2} = \abs{\phi_{1} - \phi_{2}} > 2\pi/3$. The back-to-back azimuthal selection of the jet pairs is meant to enhance the sensitivity to lower-order ($2\rightarrow 2$) partonic processes. Further selections are applied to \ptave to select data that test NLO pQCD calculations with various nPDFs at different \qq values. The \ptave intervals used in the analysis are 55--75, 75--95, 95--115, 115--150, and 150--400\GeV. The last interval is denoted by `$>$150\GeV' in the figures. The \pPb results differ from the ones reported in Ref.~\cite{Chatrchyan:2014hqa} in that a lower \pt for the leading and subleading jets was used (90 \vs 120\GeV, and 20 \vs 30\GeV, respectively), and in that the present measurement is differential in \ptave (5 \vs 1 intervals).

\begin{figure}[h]
\centering
\includegraphics[width=0.45\textwidth]{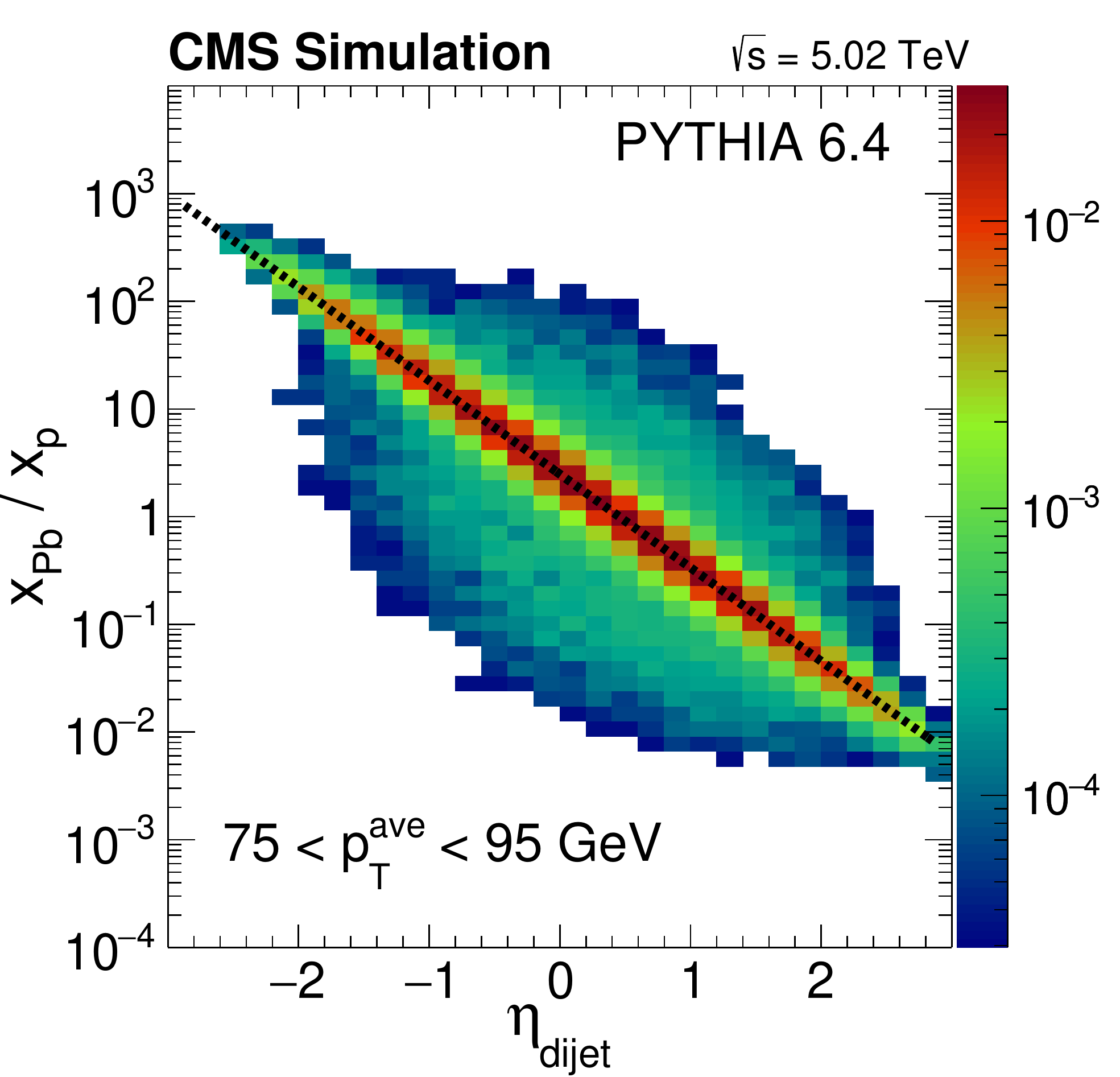}
\includegraphics[width=0.45\textwidth]{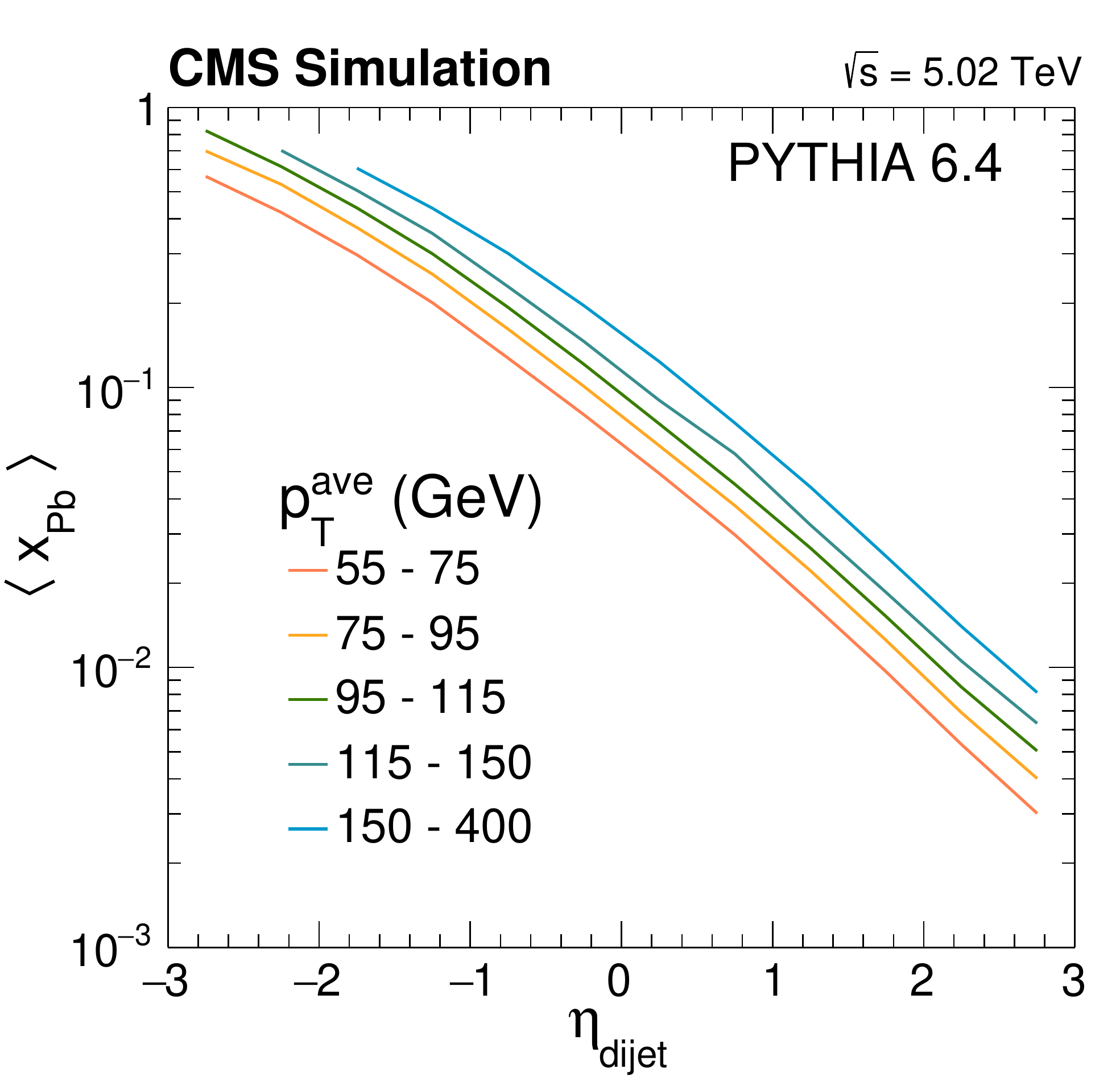}
\caption{(\cmsLeft) Correlation between $x_{\Pb}/x_{\Pp}$ and dijet pseudorapidity \etadijet. The dashed line corresponds to the expected correlation without ISR or FSR effects. (\cmsRight) Mean Bjorken $x$ of the parton from the lead ion $\left<x_{\Pb}\right>$ obtained from \PYTHIA~6 events as a function of \etadijet in different dijet \ptave intervals.}
\label{fig:xPbvsEtaDijet}
\end{figure}

In the following we discuss the relation between the kinematics of a dijet event to parton level quantities. We define $x_{\Pp}$ as the Bjorken $x$ of the parton from the nucleon going in the $+z$ direction and $x_{\Pb}$ as the Bjorken $x$ of the parton from the nucleon going in the $-z$ direction. Different regions of $x_{\Pp}$ and $x_{\Pb}$ can be chosen by selecting ranges of \etadijet. In a simple case of two partons colliding without initial-state radiation (ISR) or final-state radiation (FSR), \etadijet in the center-of-mass frame would be equal to $\frac{1}{2}\ln(x_{\Pp}/x_{\Pb})$. The effect of ISR and FSR on this correlation was studied using the {\sc pythia} event generator~\cite{bib_pythia} (version 6.423, tune Z2)~\cite{Field:2010bc}, and was found to be small, as shown in Fig.~\ref{fig:xPbvsEtaDijet} (\cmsLeft) for the $75 < \ptave < 95\GeV$ interval. The Pearson's correlation coefficient between $\ln(x_{\Pb}/x_{\Pp})$ and \etadijet is $-0.58$ in this \ptave interval. In the presence of a strong linear correlation, this coefficient would be close to $\pm1$, while in the absence of any correlation it would be closer to 0. The correlation between $\left<x_{\Pb}\right>$ and \etadijet shown in Fig.~\ref{fig:xPbvsEtaDijet} (\cmsRight) allows the identification of \etadijet intervals which are sensitive to shadowing ($\etadijet \ga 1.5$), antishadowing ($-0.5 \la \etadijet \la 1.5$), and EMC effects ($\etadijet \la -0.5$).

The systematic uncertainty related to the jet energy scale (JES) is important since the width of the \etadijet distribution decreases with increasing \ptave~\cite{Chatrchyan:2014hqa}. Studies with dijet and \GAMJET\ events~\cite{Chatrchyan:2011ds} show that the JES in data can deviate from that in simulated events by up to 2\%. To evaluate the corresponding uncertainties, the JES is shifted by $\pm 2\%$ for both \pp and \pPb data and the deviations of the observed spectra are taken as systematic uncertainties. To account for the uncertainties related to the jet energy (angular) resolution, the differences between the \etadijet spectra obtained from detector-level (\ie reconstructed) jet \pt ($\eta$) and generator-level (\ie, MC truth) jet \pt ($\eta$) with \PYTHIA for \pp and \PYTHIA events embedded in simulated \pPb underlying events (\pythiahijing) for \pPb collisions are quoted as a systematic uncertainty. To model the \pPb UE, minimum bias \pPb events are simulated with the {\HIJING} event generator~\cite{Wang:1991hta}, version 1.383~\cite{Gyulassy:1994ew}. The parameters used in the {\HIJING} simulation are tuned to reproduce the total particle multiplicities and charged-hadron spectra, and to approximate the UE fluctuations seen in data.

Other sources of uncertainties are the effects of the UE and pileup events in \pPb collisions. Combinatorial jets coming from nucleon-nucleon collisions that happen simultaneously with the hard-scattering of interest are studied using \pythiahijing simulations. The effect of the remaining pileup events in \pPb collisions is evaluated by comparing the results with and without the pileup filter. Those uncertainties are negligible compared to other sources. The total systematic uncertainties in \etadijet and in the ratios of the \pPb and \pp spectra are evaluated by summing in quadrature over the contributions from the above sources. In the $\pPb/\pp$ \etadijet ratio measurements, the uncertainties due to the JES, jet energy resolution, and jet angular resolution are partially canceled and the total systematic uncertainties are between 2 and 20\%, increasing from high- to low-\ptave values, and towards higher $\abs{\etadijet}$ values.

\begin{figure*}[htb]
\centering
\includegraphics[width=\cmsFigTwoWidth]{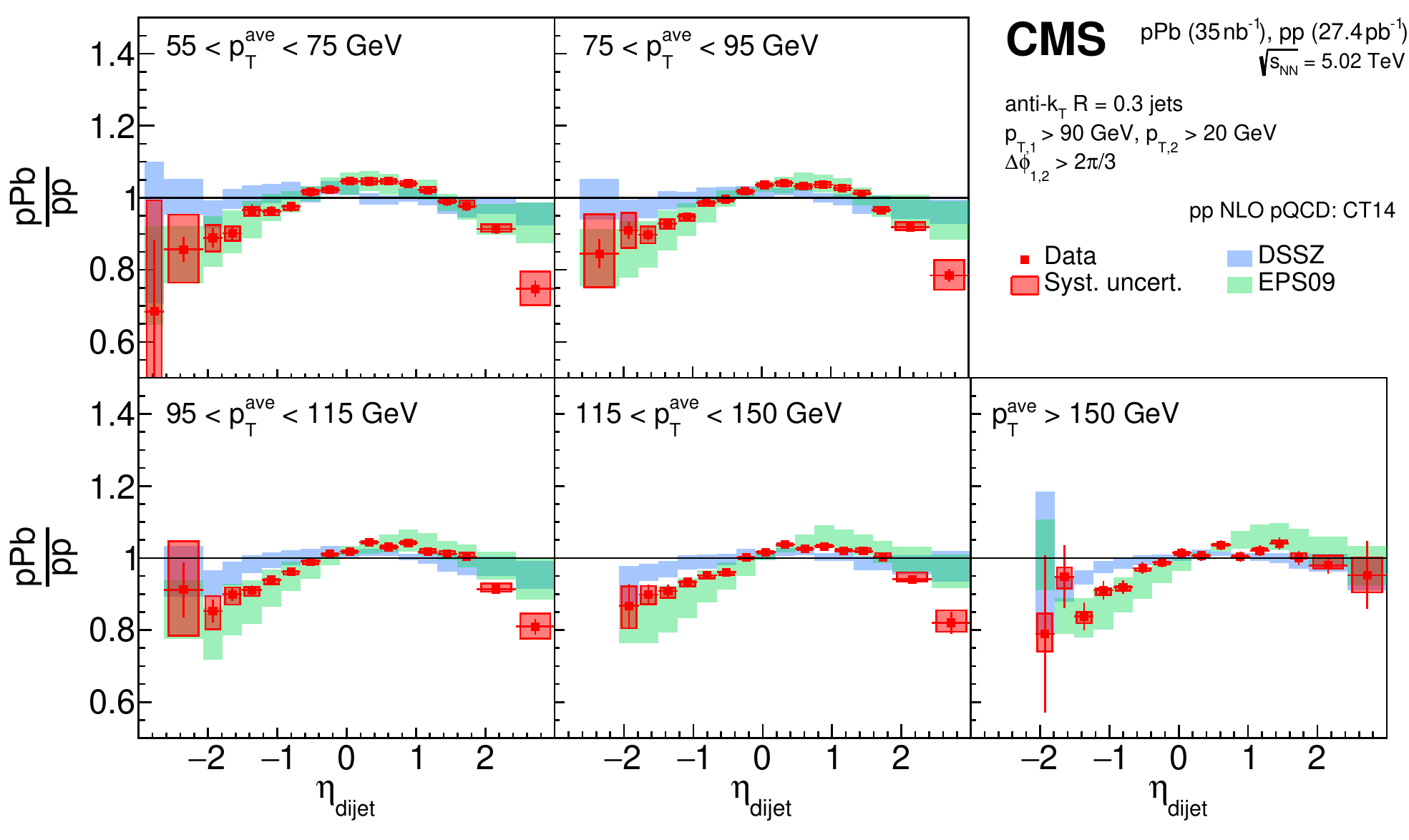}
\caption{Ratio of \pPb to \pp \etadijet spectra compared to NLO pQCD calculations with DSSZ~\cite{deFlorian:2011fp} and EPS09~\cite{Eskola:2009uj} nPDFs, using CT14~\cite{Dulat:2015mca} as the baseline nucleon PDF. Red boxes indicate systematic uncertainties in data and the height of the NLO pQCD calculation boxes represent the nPDF uncertainties.}
\label{fig:pPbppRat1}
\end{figure*}

\begin{figure}[htb]
\centering
\includegraphics[width=\cmsFigOneWidth]{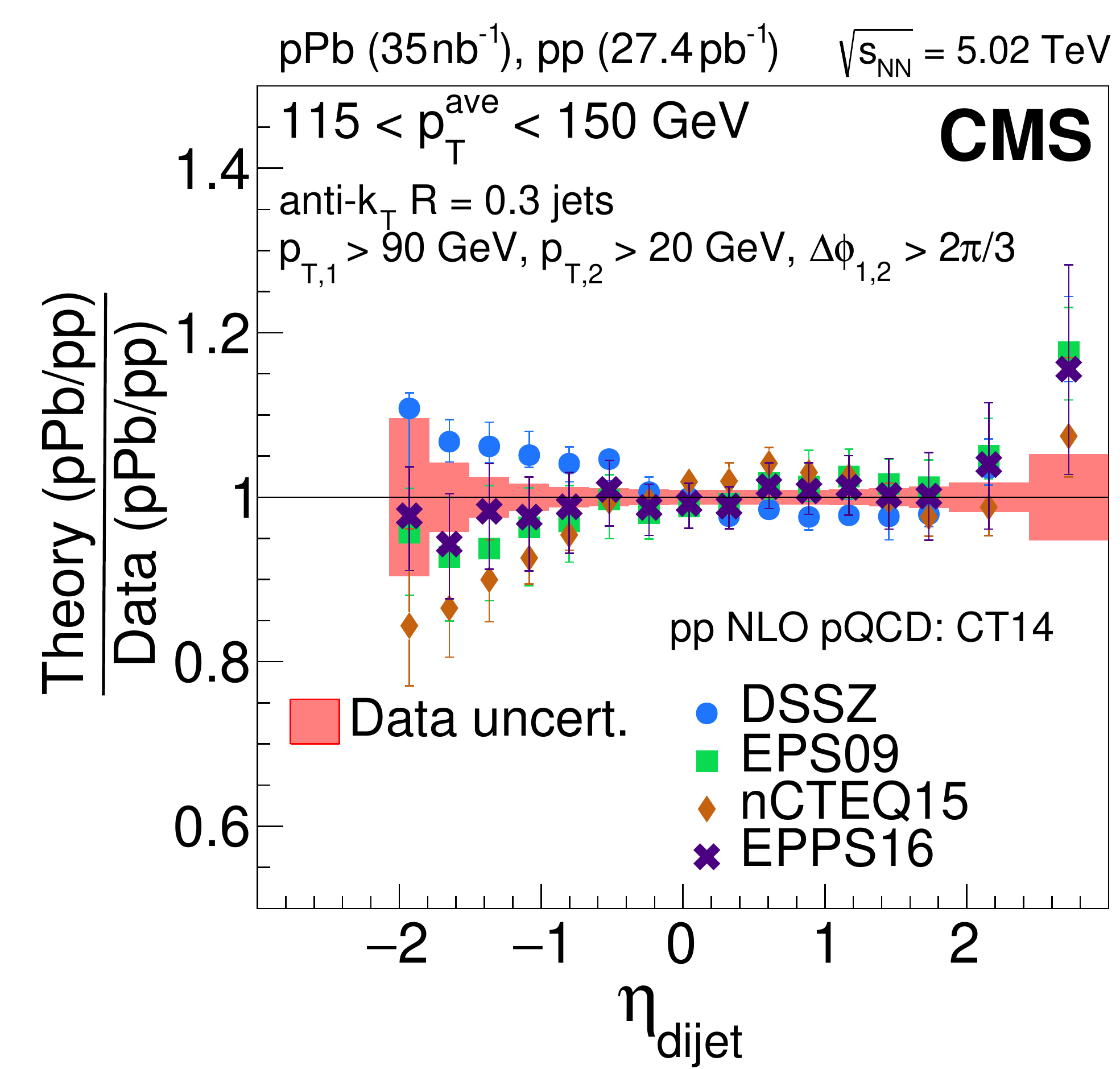}
\caption{Ratio of theory to data, for the ratio of the \pPb to \pp \etadijet spectra for $115<\ptave<150$\GeV. Theory points are from the NLO pQCD calculations of DSSZ~\cite{deFlorian:2011fp}, EPS09~\cite{Eskola:2009uj}, nCTEQ15~\cite{Kovarik:2015cma}, and EPPS16~\cite{Eskola:2016oht} nPDFs, using CT14~\cite{Dulat:2015mca} as the baseline PDF. Red boxes indicate the total (statistical and systematic) uncertainties in data, and the error bars on the points represent the nPDF uncertainties.}
\label{fig:pPbppRat2}
\end{figure}

The measured \etadijet spectra in \pp collisions, shifted to match the range of the \pPb data as described previously, and the corresponding \pPb results, are available in \suppMaterial, which includes Refs.~\cite{deFlorian:2011fp, Eskola:2009uj, Kovarik:2015cma,Dulat:2015mca,Harland-Lang:2014zoa}. In order to construct an observable that is relatively insensitive to the \pp PDF calculation~\cite{Armesto:2015lrg}, the ratios of the \pPb and \pp reference distributions, individually normalized to one, are chosen. This assumption was tested by comparing the NLO spectra ratio in pQCD calculations with CT14 and MMHT14 PDFs~\cite{Nagy:2003tz}. The shape of the ratios of the \pPb and \pp distributions in data are compared with NLO pQCD calculations based on the EPS09 and DSSZ nPDFs in Fig.~\ref{fig:pPbppRat1}. In addition, in Fig.~\ref{fig:pPbppRat2}, the ratio of the $\pPb/\pp$ \etadijet distributions in data is compared also to that from calculations based on the nCTEQ15 and EPPS16 nPDFs, for $115 < \ptave < 150\GeV$. The ratios of \pPb and \pp data are seen to deviate significantly from unity in the small (EMC) and large (shadowing) \etadijet regions. In the interval $\etadijet<-1$, which is sensitive to the gluon EMC effect, NLO pQCD calculations with EPS09 nPDF match the data at the edge of the theoretical uncertainty, while the calculations with DSSZ nPDF, where no gluon EMC effect is present in the global fit, overpredict the data.

The differences between data and the various NLO pQCD calculations with nPDFs in the interval $\etadijet<-1$ are quantified by comparing the two distributions with a $\chi^2$-test, taking into account the point-to-point correlations from the nPDFs. The uncertainties from data are taken to be uncorrelated point-to-point. For $115 < \ptave < 150\GeV$, the $p$-values from the test are 0.19, $<10^{-8}$, and $<10^{-8}$ for the EPS09, DSSZ, and nCTEQ15 nPDFs, respectively. Across the full \ptave range, the $p$-values for EPS09 range from 0.19 to 0.95, whereas the $p$-values for the DSSZ and nCTEQ15 nPDFs are never larger than 0.015. This shows that, with a $p$-value cutoff of 0.05, the data are incompatible with the DSSZ and nCTEQ15 nPDFs, but not incompatible with EPS09. This supports the interpretation of the RHIC pion data by the EPS09 nPDF, in which the modification of the pion spectra gives rise to the gluon EMC effect. The data also show smaller shadowing, antishadowing, and EMC effects than what is implemented in the nCTEQ15 PDF set. The results are consistent with EPPS16 with relaxed constraints (\eg, more free parameters, increased error tolerance) on the nuclear PDF parametrization, which results in larger PDF uncertainties~\cite{Eskola:2016oht}. The conclusions obtained from different \ptave intervals are similar, which provide important tests of the nuclear PDF at various \qq values. The significantly smaller experimental uncertainties, in most of the \ptave and \etadijet intervals probed, as compared to the uncertainties of calculations using NLO PDF, indicate that the present data, when included in the calculations of the next generation nPDFs, will result in an improved description of the gluon nPDF.

In summary, measurements of the dijet pseudorapidity (\etadijet) in different average transverse momentum (\ptave) intervals in \pPb and \pp collisions at a nucleon-nucleon center-of-mass energy $\sqrtsNN = 5.02\TeV$ are reported. Ratios of the \pPb and \pp \etadijet spectra using the \pp reference data are also reported. Significant modifications of the \etadijet distributions are observed in \pPb data compared to the \pp reference in all \ptave intervals, which show shadowing, antishadowing, and EMC effects in nuclear parton distribution functions. The ratios of the \pPb and \pp distributions are compared to next-to-leading order calculations with nucleon and nPDFs. Based on these comparisons, the results present the first evidence that the gluon PDF at large Bjorken $x$ in lead ions is strongly suppressed with respect to that in unbound nucleons. The data are incompatible with predictions using nucleon PDFs or using nPDFs without large-$x$ gluon suppression. Based on a statistical analysis, the EPS09 nPDF provides the best overall agreement with the data. This data can be used to place strong constraints on the next-generation of nPDFs, which are crucial for the understanding of high \pt and high mass particle production at collider energies.

\begin{acknowledgments}
We thank Nestor Armesto, Paukkunen Hannu, Pia Zurita, Petja Paakkinen and Carlos Salgado for the useful discussions and the NLO pQCD calculations with various nPDFs. We congratulate our colleagues in the CERN accelerator departments for the excellent performance of the LHC and thank the technical and administrative staffs at CERN and at other CMS institutes for their contributions to the success of the CMS effort. In addition, we gratefully acknowledge the computing centers and personnel of the Worldwide LHC Computing Grid for delivering so effectively the computing infrastructure essential to our analyses. Finally, we acknowledge the enduring support for the construction and operation of the LHC and the CMS detector provided by the following funding agencies: BMWFW and FWF (Austria); FNRS and FWO (Belgium); CNPq, CAPES, FAPERJ, and FAPESP (Brazil); MES (Bulgaria); CERN; CAS, MoST, and NSFC (China); COLCIENCIAS (Colombia); MSES and CSF (Croatia); RPF (Cyprus); SENESCYT (Ecuador); MoER, ERC IUT, and ERDF (Estonia); Academy of Finland, MEC, and HIP (Finland); CEA and CNRS/IN2P3 (France); BMBF, DFG, and HGF (Germany); GSRT (Greece); NKFIA (Hungary); DAE and DST (India); IPM (Iran); SFI (Ireland); INFN (Italy); MSIP and NRF (Republic of Korea); LAS (Lithuania); MOE and UM (Malaysia); BUAP, CINVESTAV, CONACYT, LNS, SEP, and UASLP-FAI (Mexico); MBIE (New Zealand); PAEC (Pakistan); MSHE and NSC (Poland); FCT (Portugal); JINR (Dubna); MON, RosAtom, RAS and RFBR (Russia); MESTD (Serbia); SEIDI, CPAN, PCTI and FEDER (Spain); Swiss Funding Agencies (Switzerland); MST (Taipei); ThEPCenter, IPST, STAR, and NSTDA (Thailand); TUBITAK and TAEK (Turkey); NASU and SFFR (Ukraine); STFC (United Kingdom); DOE and NSF (USA).
\end{acknowledgments}

\bibliography{auto_generated}

\ifthenelse{\boolean{cms@external}}{}{
\clearpage
\numberwithin{figure}{section}
\appendix
\section{\texorpdfstring{Supplemental information: Measured $\eta_\text{dijet}$ spectra in \pPb collisions}{Supplemental information: Measured eta[dijet] spectra in pPb collisions}\label{app:suppMat}}
\begin{figure*}[htb]
\centering
\includegraphics[width=0.95\textwidth]{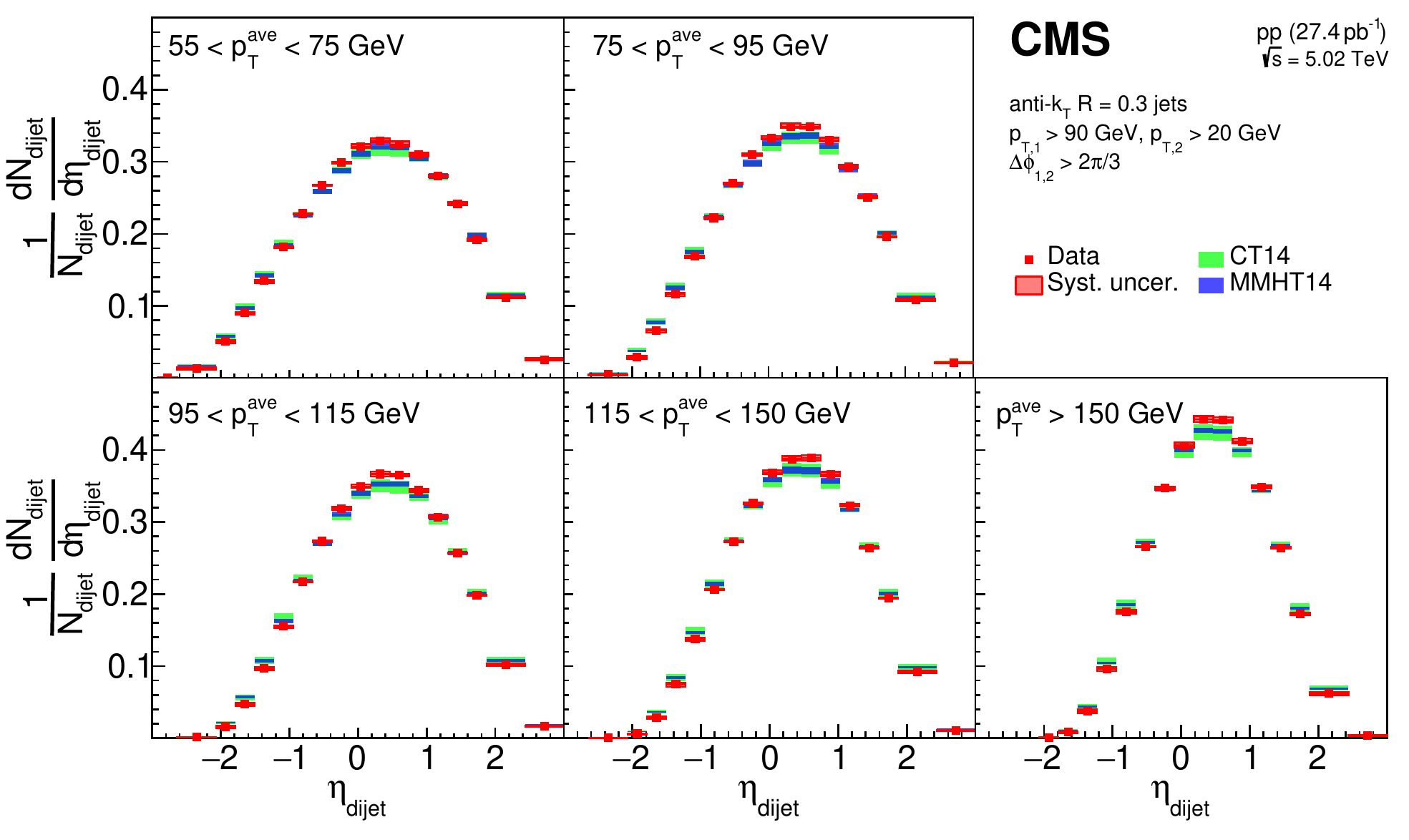}
\caption{The measured \pp dijet pseudorapidity spectra, shifted to match the \etadijet range of the \pPb collisions, in bins of \ptave, overlaid with NLO pQCD calculations with the CT14~\cite{Dulat:2015mca} and MMHT14~\cite{Harland-Lang:2014zoa} nucleon PDFs. The red boxes indicate systematic uncertainties in data and the height of the NLO pQCD calculation boxes represent the PDF uncertainties.}
\label{fig:ppDists}
\end{figure*}

\begin{figure*}[th!]
\begin{center}
\includegraphics[width=0.95\textwidth]{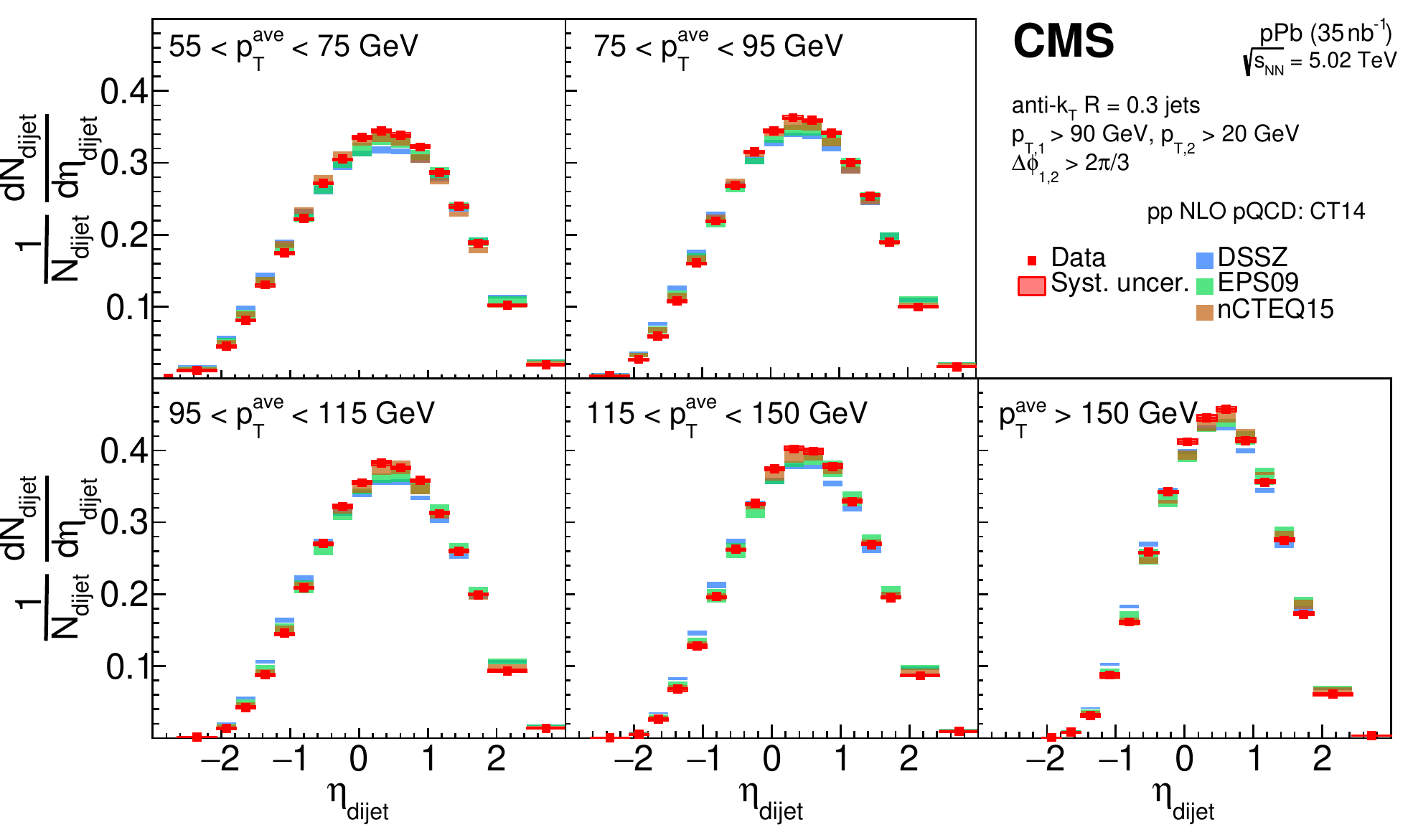}
\caption{The measured pPb dijet pseudorapidity spectra in bins of \ptave, overlaid with the NLO pQCD calculations of DSSZ~\cite{deFlorian:2011fp}, EPS09~\cite{Eskola:2009uj}, and nCTEQ15~\cite{Kovarik:2015cma} nPDFs, using the CT14~\cite{Dulat:2015mca} as the baseline PDF. The red boxes indicate systematic uncertainties in data and the height of the NLO pQCD calculation boxes represent the nPDF uncertainties.}
\label{fig:paDists_ppCT14}
\end{center}
\end{figure*}

\begin{figure*}[hb!]
\begin{center}
\includegraphics[width=0.95\textwidth]{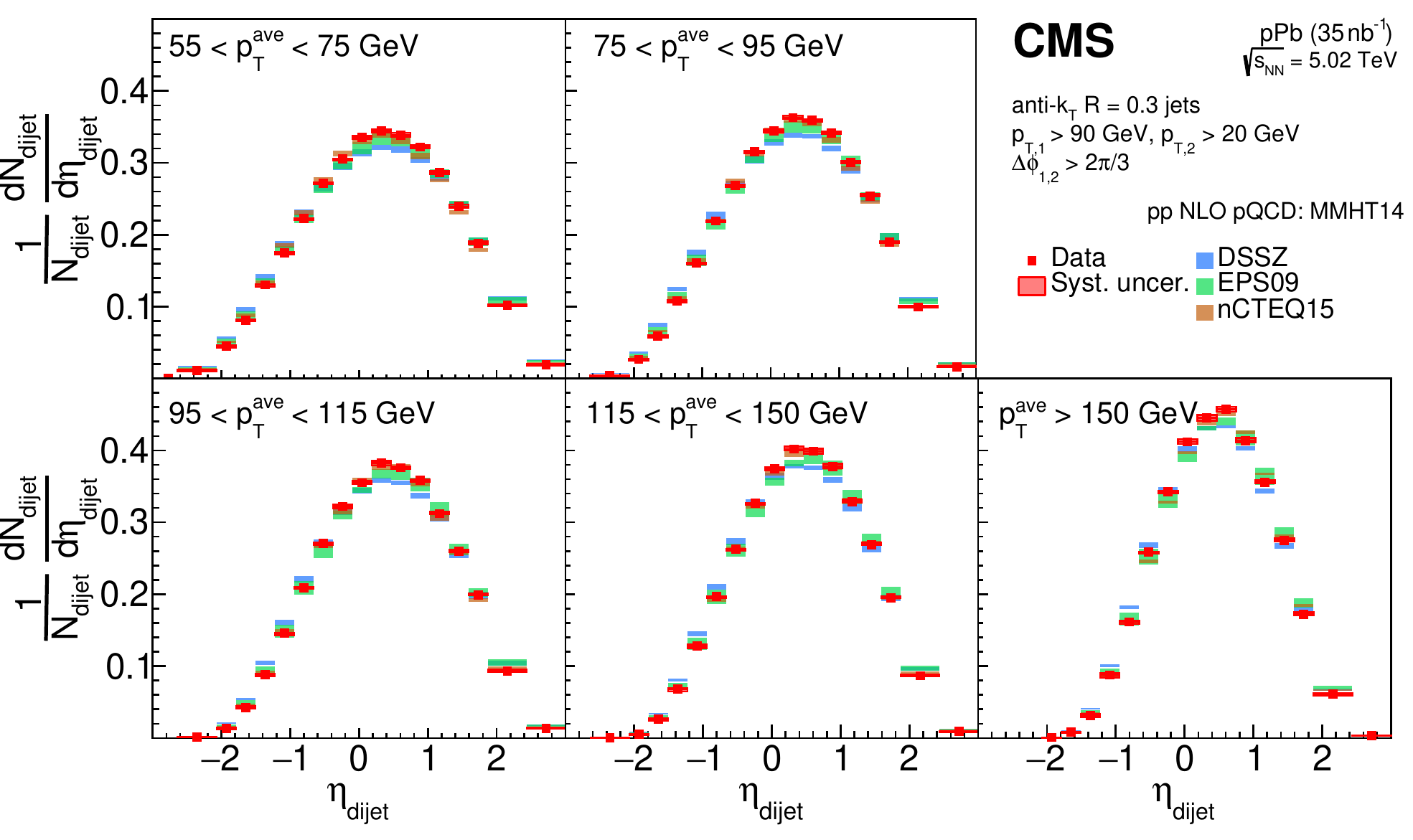}
\caption{The measured pPb dijet pseudorapidity spectra in bins of \ptave, overlaid with the NLO pQCD calculations of DSSZ~\cite{deFlorian:2011fp}, EPS09~\cite{Eskola:2009uj}, and nCTEQ15~\cite{Kovarik:2015cma} nPDFs, using the MMHT14~\cite{Harland-Lang:2014zoa} as the baseline PDF. The red boxes indicate systematic uncertainties in data and the height of the NLO pQCD calculation boxes represent the nPDF uncertainties.}
\label{fig:paDists_ppMMHT14}
\end{center}
\end{figure*}

}
\cleardoublepage \section{The CMS Collaboration \label{app:collab}}\begin{sloppypar}\hyphenpenalty=5000\widowpenalty=500\clubpenalty=5000\vskip\cmsinstskip
\textbf{Yerevan Physics Institute, Yerevan, Armenia}\\*[0pt]
A.M.~Sirunyan, A.~Tumasyan
\vskip\cmsinstskip
\textbf{Institut f\"{u}r Hochenergiephysik, Wien, Austria}\\*[0pt]
W.~Adam, F.~Ambrogi, E.~Asilar, T.~Bergauer, J.~Brandstetter, M.~Dragicevic, J.~Er\"{o}, A.~Escalante~Del~Valle, M.~Flechl, R.~Fr\"{u}hwirth\cmsAuthorMark{1}, V.M.~Ghete, J.~Hrubec, M.~Jeitler\cmsAuthorMark{1}, N.~Krammer, I.~Kr\"{a}tschmer, D.~Liko, T.~Madlener, I.~Mikulec, N.~Rad, H.~Rohringer, J.~Schieck\cmsAuthorMark{1}, R.~Sch\"{o}fbeck, M.~Spanring, D.~Spitzbart, A.~Taurok, W.~Waltenberger, J.~Wittmann, C.-E.~Wulz\cmsAuthorMark{1}, M.~Zarucki
% ...
\vskip\cmsinstskip
% ...
1:  Also at Vienna University of Technology, Vienna, Austria\\
% ...
\end{sloppypar}
\end{document}